*Article*

# Experimental Study of AM and PM Noise in Cascaded Amplifiers

Inari Badillo * and Joaquín Portilla

Department of Electricity and Electronics, University of the Basque Country UPV/EHU, 48940 Leioa, Spain; joaquin.portilla@ehu.eus
* Correspondence: inari.badillo@ehu.eus

**Abstract:** An experimental study of amplitude modulation (AM) and phase modulation (PM) noise spectra in cascaded amplifiers was carried out as a function of the number of amplification stages and the input power. Flicker and white noise contributions were determined, as well as effective noise figure (NF) from AM and PM noise spectra from small-signal to large-signal regimes. Simultaneous measurements of AM and PM noise were performed, and associated correlation was measured as a function of the offset frequency from the carrier. Measurements exhibited, in general, quite low AM–PM correlation levels both in the flicker and white noise parts of the spectrum. In some particular amplifier configurations, however, measurements showed some peaks in the correlation at some specific input power levels in the transition zone, from a quasi-linear to strong compression. The results show that the effective noise figure decreases with the number of stages for a given carrier output power level.

**Keywords:** non-linear noise; cascaded amplifiers; AM noise; PM noise; flicker noise; white noise; noise figure

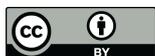

## 1. Introduction

Amplitude modulation (AM) and phase modulation (PM) noises have always been important concerns in radio communication, radar, radiometry, or RF and microwave particle acceleration, among other applications (see, for instance, [1–10]). The study of AM and PM noise is a useful diagnostic tool for device technology optimization and aging studies [11].

The AM and PM noise power produced in one amplifier depend on the carrier level, both in the near-carrier and white regions of the carrier noise spectra [1–10]. For a given carrier frequency and power, different levels of flicker and white AM and PM noise can be obtained depending on device bias or matching conditions [12,13]. Measurement and simulation results have shown that AM and PM noise contributions are identical just in the white region of the noise spectra and only in the case of the amplifier working under linear conditions [12,13]. If this is the case, the noise figure (NF) coming from AM or PM noise spectra is equal to the standard NF, obtained from different characterization techniques such as Y-factor [14] or cold source [15]. As carrier power grows, measurements and simulations have shown that both flicker and white noise contributions are not the same for the AM and PM cases [12,13]. In particular, flicker AM and PM noise power levels are different even when they are measured under a small signal regime. This means that the nonlinear mechanisms producing converted AM and PM noises are specific to each kind of noise. As a consequence, the effective NF can exhibit significant degradation with input power, and it must be carefully evaluated by taking into account not only PM white noise contribution, but also AM white noise, as demonstrated in [12].

In this paper, an experimental study of AM and PM noise in amplifiers working in small- and large-signal regimes is devoted to cascaded amplifier configurations. The AM and PM usual figures of merit refer to the noise produced by the amplifier to the carrier power, and are shown to be poorer by increasing the number of stages when compared either for a given input or output carrier power level. This is more noticeable as the amplifiers are driven into compression. Nevertheless, for a given output power, effective NF can significantly be improved by adding some stages. Experimental results are reported and discussed in the paper. Moreover, the study was also extended to the evaluation of the AM–PM noise cross-correlation by measurement of the coherence function [16], defined in terms of noise power spectral densities, as a function of the offset from the carrier frequency, from low-frequency noise to white noise contributions to the carrier noise. These kinds of measurements were obtained using a system based on an I-Q receiver scheme, coupled to data acquisition and processing techniques and applying specific setting and calibration techniques. The paper is organized into four sections plus the final conclusions. In Section 2, the measurement setup is described. Section 3 is devoted to summarizing the experiment results and discussing them. Experimental results are discussed in Section 4 with the help of an analytical simplified model of the amplifier. The noise mechanisms contributing to the carrier noise in RF and microwave cascade amplifiers are considered.

## 2. Measurement Setup

The measurement system is based on a I-Q receiver architecture, including 16-bit precision ADCs and a self-calibration. This system is coupled to data acquisition and processing techniques allowing for the simultaneously measuring of AM and PM noise of the device under test (DUT), as well as the correlation between both contributions to carrier noise. As shown in Figure 1, the source CW signal is divided into two branches, one that is connected to the RF branch of the receiver and the other to the local oscillator (LO) branch. The DUT, together with a passive phase shifter, is placed into the LO branch, between the power splitter and the LO input port. The phase shifter must be adjusted to obtain proper settings for the measurement system as explained below.

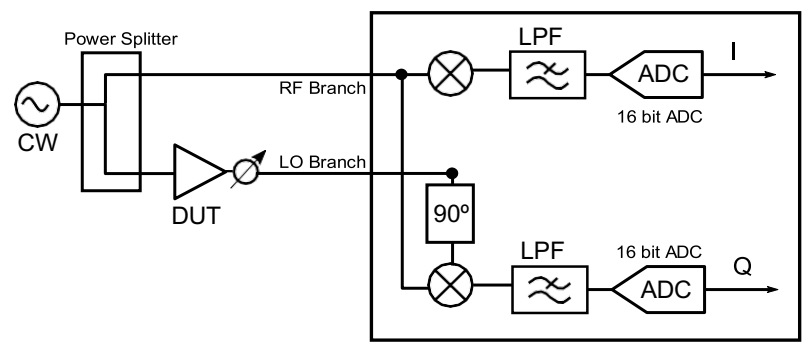

**Figure 1.** Schematic of the experimental setup for the measurements.

The system is settled in order to measure the DUT AM noise from the I branch output, eliminating PM noise contributions from the source and the DUT itself, and to obtain the PM noise produced by the DUT from the Q branch output, eliminating in this case the DUT AM noise contribution, as well as the AM and PM noise introduced by the source. With the appropriate setting and calibration procedure [17], this setup allows us to study the evolution of both AM and PM noise produced by the DUT in the presence of carrier signals under linear and nonlinear regimes, as well as their correlation through the measurement of the well-known coherence function, which provides the normalized correlation as a function of the offset from the carrier frequency. In this paper, [17], the main contributors to measurement errors, and the methodology to obtain maximum achievable resolution (obtaining an uncertainty of $\pm 0.15$ dB) using this kind of measurement setup, are discussed in detail with illustrative examples.

Ideally, when the phase difference of the two signals coming into the Q-branch double-balanced mixer is exactly 90°, it acts like a phase detector. The desired phase difference can be obtained simply by tuning the phase shifter. The variations between the phase fluctuations in both the signals at the mixer input appear as voltage changes at the mixer output [1–3]. The PM noise from the source is then eliminated. The double-balanced mixer employed in each demodulator branch notably reduces the AM noise coming from the system blocks, at the same time avoiding AM to PM conversion in their own mixing process [1–3].

In the case of the AM measurement, both the signals coming into the I-branch mixer input are ideally in phase. Therefore, the double-balanced mixer acts as an amplitude detector in this case, because the mixer output voltage is proportional to the amplitude variations of the signals coming into the mixer inputs. The suppression of the phase fluctuations can be up to 90 dB [3] using this configuration.

## 3. Results

AM and PM noise power spectra measurements at different carrier levels, from linear to strong compression operation and for a different number of amplification stages, were obtained from the experimental set up described in the previous section. The results reported in this paper correspond to a family of cascade amplifiers built up using up to four identical amplification stages. Each stage shows about 16 dB of linear gain, an NF around 5 dB and a good matching at the working frequency of 2 GHz. AM and PM noise power spectra measurements corresponding to the one-stage amplifier configuration are shown in Figures 2 and 3, respectively. In fact, the one-stage amplifier is shown to be the more challenging one to measure due to its smaller gain and lower added noise power levels.

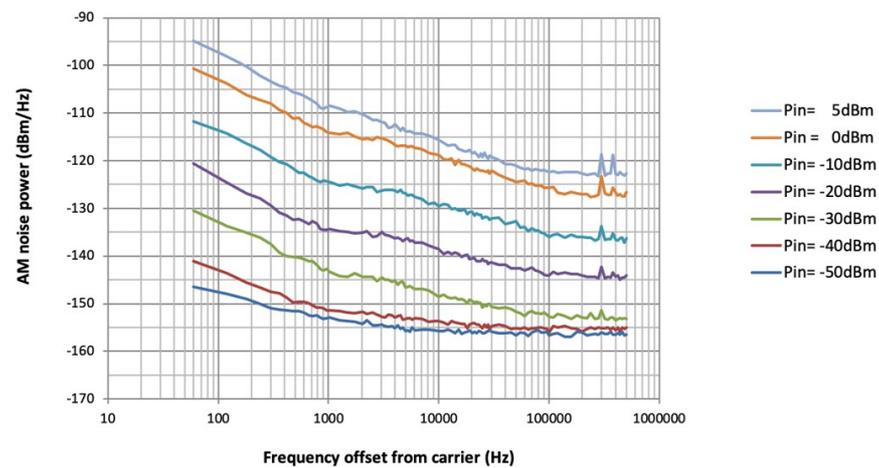

**Figure 2.** AM noise power spectra of one-stage amplifier with the carrier input power as a parameter.

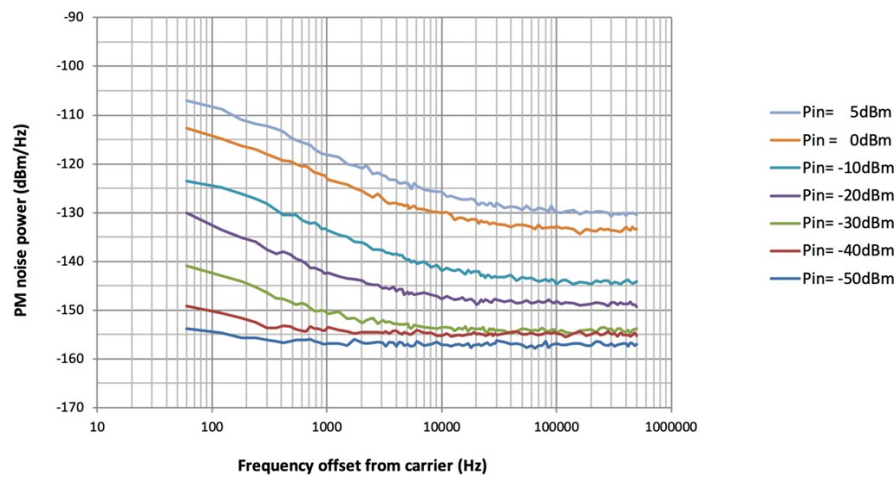

**Figure 3.** PM noise power spectra of one-stage amplifier with the carrier input power as a parameter.

The simultaneous measure of the AM and PM noise spectra allows us to determine the AM to PM correlation by the evaluation of the coherence function, which is defined as a function of the offset frequency from the carrier. The coherence function ranges from 0 for no correlated signals, to 1 in the case of completely correlated signals. The coherence function values obtained for white and flicker noise as a function of input power, and for a different number of amplification stages, are reported in Figures 4 and 5. No relevant correlation was obtained between AM and PM noise produced at different carrier power conditions in those particular amplifiers.

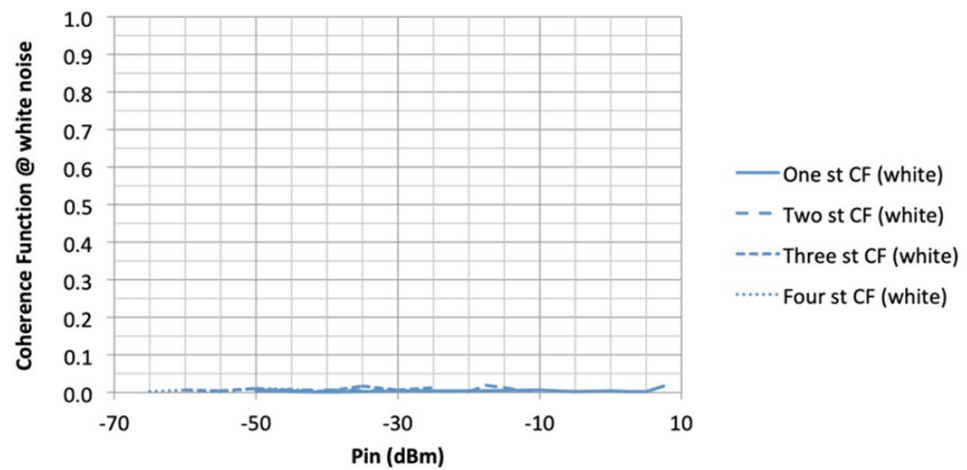

**Figure 4.** AM-to-PM coherence function in the white noise region versus carrier input power, measured for one- to four-stage amplifiers.

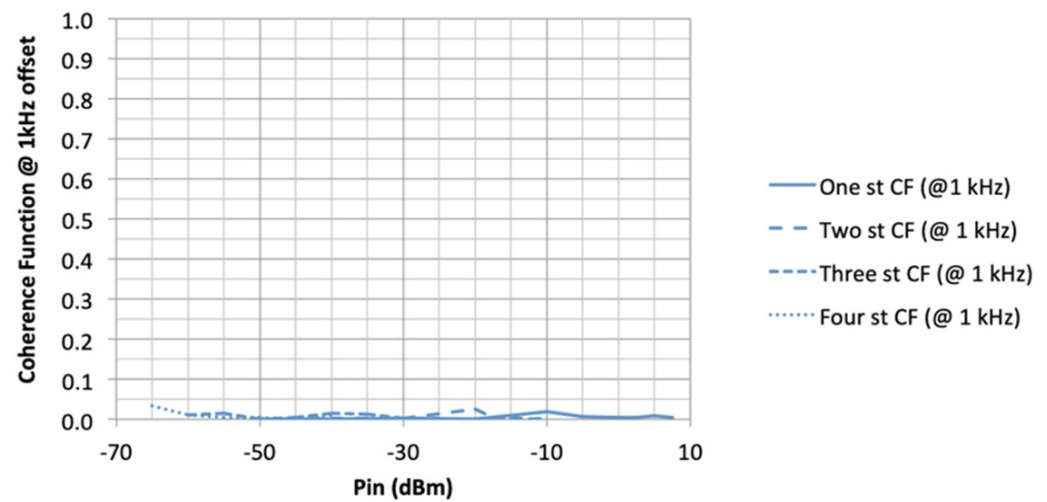

**Figure 5.** AM-to-PM coherence function, measured at 1 KHz offset frequency, versus carrier input power for one- to four-stage amplifiers.

AM and PM noise performances as a function of input power and for 1–4 amplification stages are shown in terms of single-sideband noise power, with respect to carrier power [18], in Figures 6 and 7 for the white portion of the spectra and for flicker noise at the 1 kHz offset, respectively.

It can be noticed that, as the number of stages is increased, the AM and PM noise performance worsens but is still comparable under small-signal conditions. The noise power produced by the amplifiers remains quite independent of the carrier power level, and this is the reason why the AM and PM performances, which are referred to by this carrier power, initially decrease linearly with input power. On the other hand, the amplifiers are driven into compression at lower levels of carrier input power as the number of stages increases. Therefore, the noise performance deviates beforehand from their initial linear behavior, thus indicating that the noise power grows significantly with carrier power in such conditions.

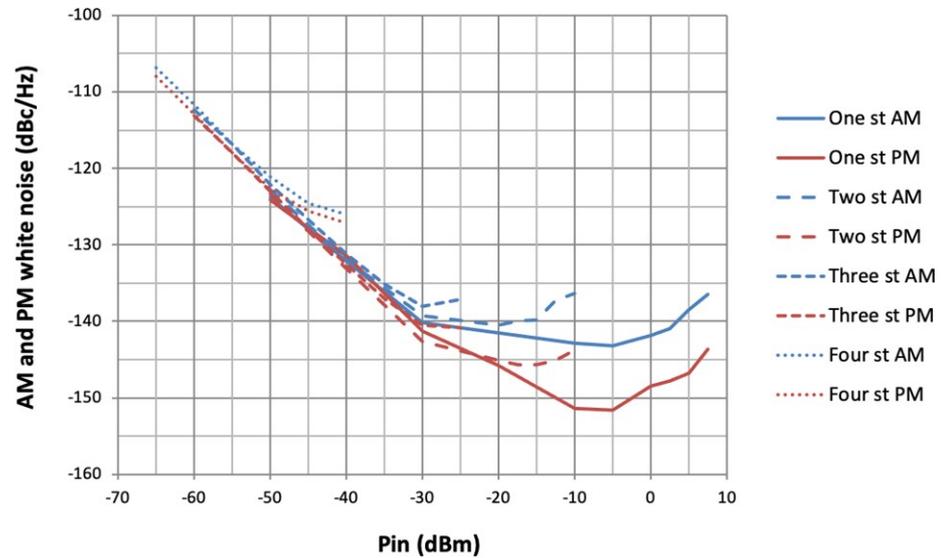

**Figure 6.** Measured AM and PM white noise versus carrier input power, produced in one- to four-stage amplifiers.

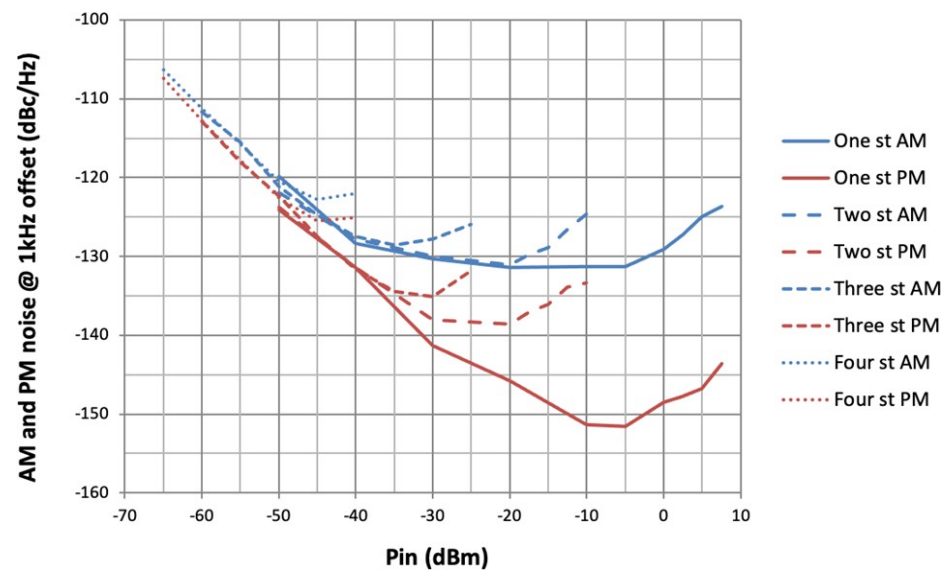

**Figure 7.** AM and PM noise measured at 1 KHz offset frequency versus carrier input power, corresponding to one- to four-stage amplifiers.

It is also interesting to observe the noise behavior in terms of effective NF (see Figure 8), which is obtained from AM and PM noise in the white region of the noise spectra around the carrier [12]. It can be observed that effective NF shows quite similar values at a low input power, according to the Friis expression [19]. Then, it rises quickly with input power as the number of stages is increased but the degradation of effective NF is lower for the amplifier with more stages, even if all the amplifiers are driven up to similar gain compression levels in the experiments.

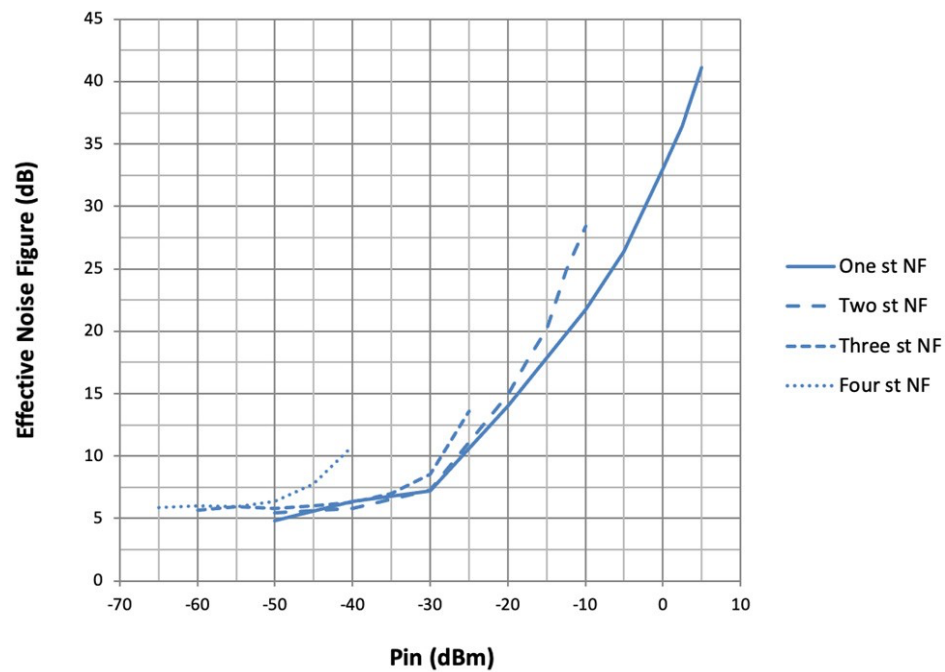

**Figure 8.** Measured effective NF versus carrier input power, associated with one- to four-stage amplifiers.

## 4. Discussion

In most situations, white noise commonly is the main noise concern under small-signal amplification. In cascaded amplifiers, the white noise added by one stage is subsequently amplified by the stages afterwards, which in turn adds its own noise contributions [19]. The NF of the overall amplifier increases slightly with the number of stages but remains comparable provided that they have enough gain. This is what we can observe in Figure 8 at low input power levels.

Since amplifiers are driven into compression, the effective NF experiments showed that effective NFs rise with carrier input power, and this becomes faster when adding amplification stages due to boosted non-linear effects in the amplifying chain, producing an augmentation of the white noise contributions with respect to the small-signal regime. Note that the results in Figure 8 correspond to amplifiers built up by using a different number of identical stages, but all of these amplifiers have been driven to similar gain compression levels. Effective NF degradation in absolute terms is significantly more important in the amplifiers with lower gain (fewer stages). A practical consequence is that, for a given output power level, the effective noise figure can be significantly improved by adding some amplification stages, except under small-signal conditions for which it remains essentially constant or, more precisely, slightly worse, as shown in Figure 9. It can be argued that achieving an identical output power implies that the noise power contribution of the last stage is identical for the all of the amplifiers, independent of their number of stages. Nevertheless, by adding stages, their contributions are minimized by the gain of previous stages that, moreover, contributes with a lower amount of noise power provided that they are working at more reduced power driving.

More generally, whether the noise is flicker or white in nature, it is assumed in practice that the small power of noise produces a small perturbation that produces a low-level modulation of the carrier amplitude and phase. The resulting carrier noise is usually analyzed by considering that it constitutes a linear perturbation of the time-varying steady state imposed by the presence of the carrier signal. This kind of analysis approach can be performed at circuit level by using a modulation approach or the so-called conversion-matrix technique [20], already available in most commercial harmonic-balance circuit simulators. The additive and converted noise modulates the amplitude and phase of the carrier signal producing the so-called AM and PM noise sidebands. At a given offset frequency from the carrier, this process can be modeled as a small-index AM or PM modulation, in which the corresponding modulating signal is a small-amplitude pseudo-sinusoidal noise source [20–22].

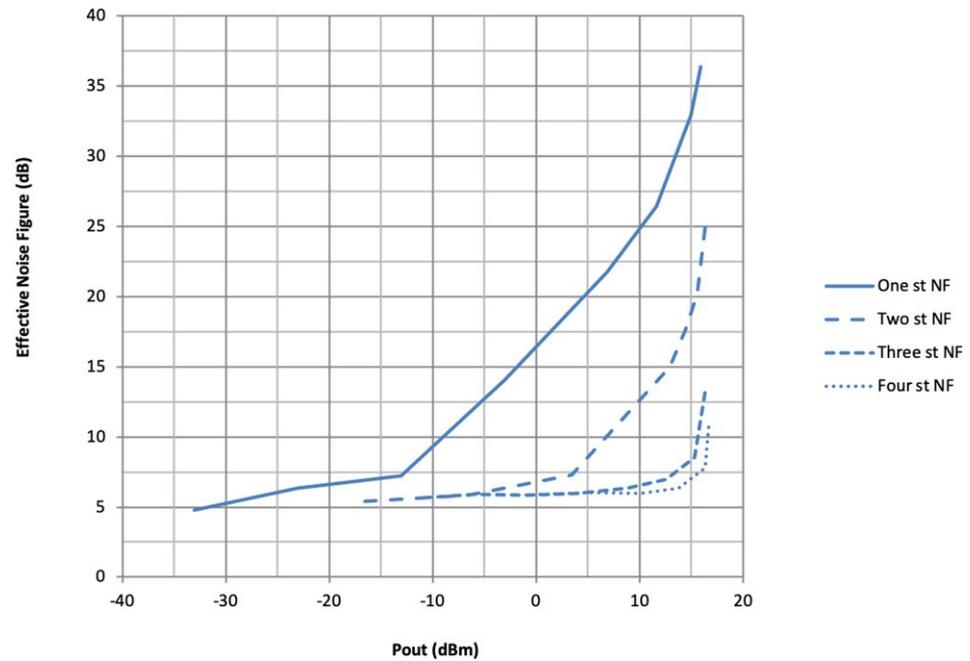

**Figure 9.** Measured effective NF versus carrier output power, associated with one- to four-stage amplifiers.

Each stage of a cascade amplification chain will amplify the incoming AM and PM small-level noise modulations produced in previous stages and will add its own contribution. In Figure 10, AM and PM noise power at the output of the amplifiers are shown as a function of the input carrier power at 1 kHz offset from the carrier frequency. These results correspond to those that were presented as a function of the carrier power in Figure 7.

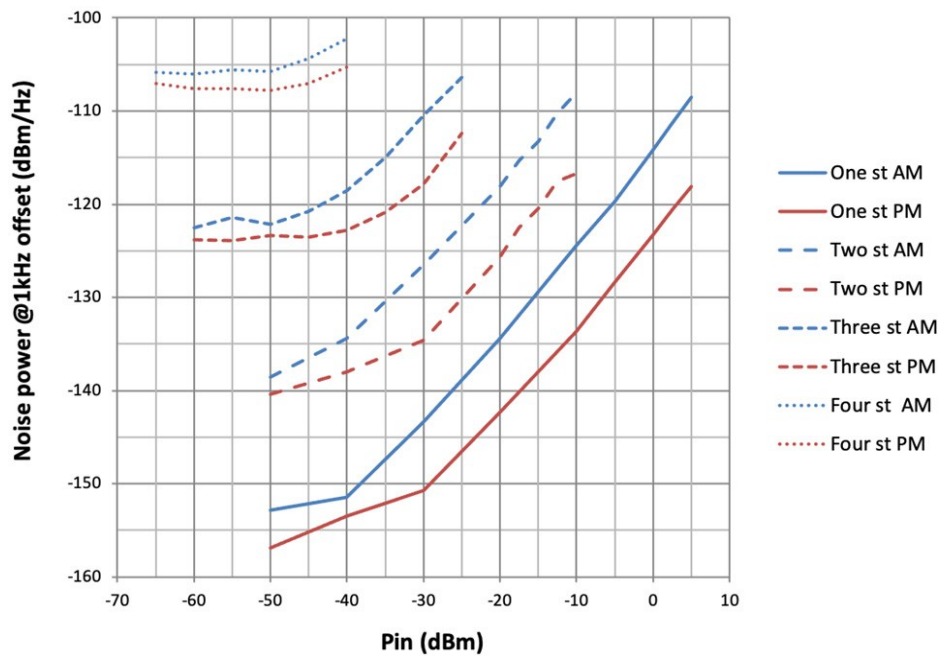

**Figure 10.** AM and PM noise power measured at 1 KHz offset frequency versus carrier input power, produced into one- to four-stage amplifiers.

From Figure 10, it can be argued that AM and PM noise produced at the first stage is the dominant contribution due to the subsequent amplification. By adding stages, the amplified noise power is the dominant contribution to overall noise, with respect to the noise produced by the individual contributions of the added stages, and this is the case from small-signal to mild non-linear regimes. At high gain compression levels, on the other hand, noise contributions are observed to grow faster with carrier power. Note that, the fact of adding stages does not produce an improvement in AM and PM noise performances, neither when compared for a given input power nor for a given output power. This is because each stage provides an equal amplification gain to the signal and the noise coming from previous stages and, moreover, adds its own noise contribution, thus resulting in increased levels of AM and PM noise as usually expressed in dBc/Hz units.

In the following, we discuss the results concerning AM to PM correlation experiments. According to the results in the previous section and experiments performed in other amplifiers, we observed that the correlation between AM and PM noises was very low, indicating that the mechanisms producing both types of noise behave as if they mostly act independent. Nevertheless, we detected specific amplifiers for which this was not the case under particular operating conditions. They correspond to one- and two-stage amplifiers made up by using another type of amplifier component. In this case, the linear gain of each amplifier stage is over 26 dB, and the NF is around 1 dB at a frequency of 2 GHz. The measured results of the AM-to-PM coherence function are presented in Figures 11 and 12, respectively, for the one-stage and two-stage amplifiers, as a function of carrier input power levels and for different offset frequencies from the carrier. Some peaks can be observed in the coherence function in both amplifiers, as well as in the flicker and white noise regions of the spectra but with more emphasis at some particular frequency offsets in the flicker noise region around a 10kHz offset. Moreover, this behavior appears at some particular carrier power levels that correspond to around the 1 dB gain compression point in each particular amplifier chain.

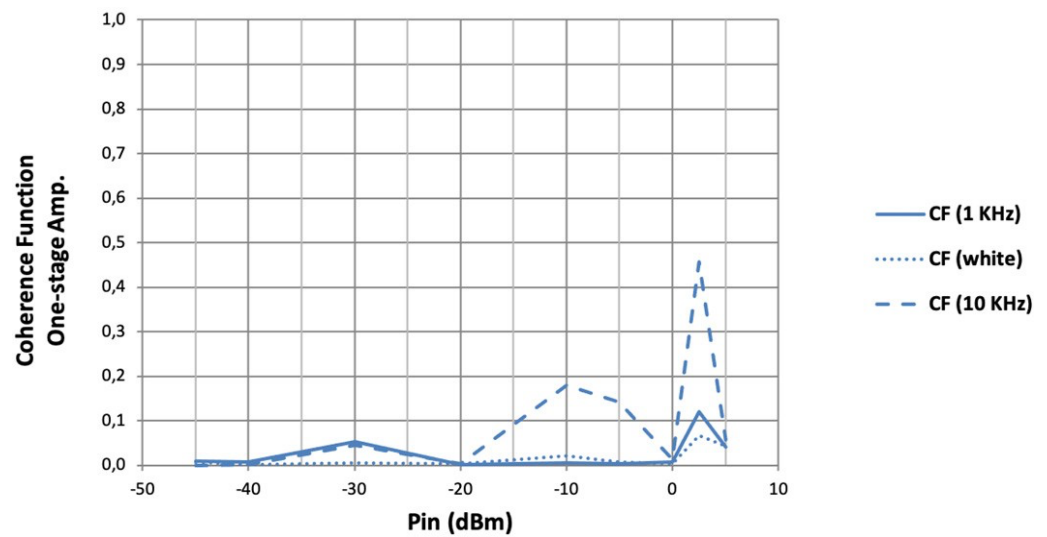

**Figure 11.** AM-to-PM coherence function versus carrier input power for the one-stage amplifier, measured at 1 KHz offset, at 10 KHz offset and in the white noise region.

In [13], it is shown that the dependence on the carrier power of AM and PM noise behavior in one stage could be linked to the derivatives of the AM–AM and AM–PM curves, evaluated in the steady-state regime forced by the particular carrier amplitude and frequency. Regarding the experimental results of the AM-to-PM correlation shown in Figures 11 and 12, the AM-to-PM derivatives were measured in both one- and two-stage amplifiers, and the results are reported in Figure 13. A noticeable similar peaking behavior is obtained in such derivatives at identical input power conditions compared to the peaks in the coherence function results of Figures 11 and 12. This fact reveals that the non-linear mechanisms behind the noise conversion, which produce an overall noise performance of the amplifier as a combined effect of different small-amplitude noise sources inside the device, simultaneously play a significant role in the input–output characteristics of the amplifier, such as the AM–PM response.

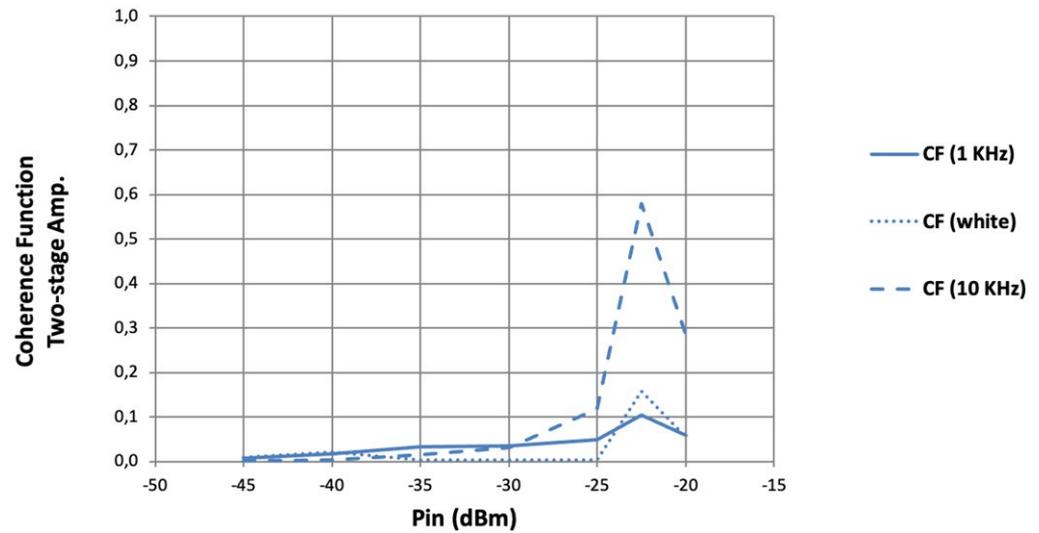

**Figure 12.** AM-to-PM coherence function versus carrier input power for the two-stage amplifier, measured at 1 KHz offset, at 10 KHz offset and in the white noise region.

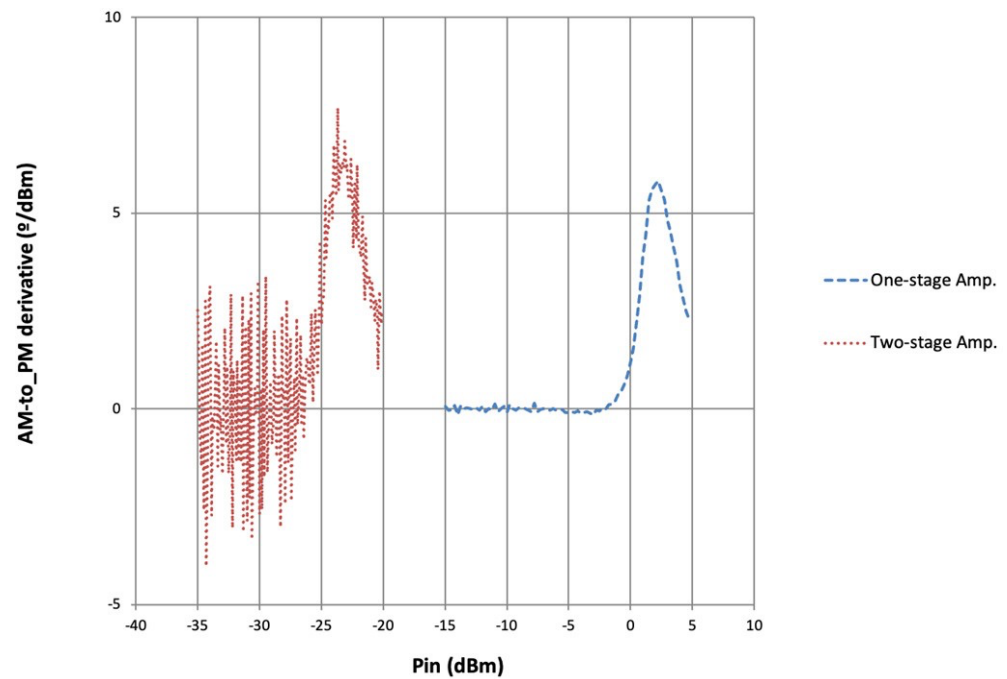

**Figure 13.** Measured AM-to-PM derivative versus carrier input power.

It is well known that the correlation is zero as far as additive white noise is concerned [22]. Due to the particular noise mechanisms that control AM and PM noise conversion around the carrier, it can be conjectured that AM-to-PM noise correlation would be negligible or low under a small-signal regime and could grow under a high compression. Moreover, this is expected to be more noticeable in any case of near-carrier noise compared to white noise due to the intrinsically non-auto-correlated nature of white noise. However, in any case, under a strong periodic excitation, noise becomes cyclo-stationary [21] and the AM-to-PM correlation could become higher.

## 5. Conclusions

An experimental study of the AM and PM noise in cascaded amplifiers working in small- and large-signal regimes was reported, including the evaluation of the AM–PM noise correlation by measurement of the coherence function, which determines the correlation as a function of the offset from the carrier frequency. Measurements were performed using a system based on an I-Q receiver scheme, coupled to data acquisition and processing methods, and by applying specific settings and calibration techniques. As the main conclusions, we reported that the effective noise figure, which has to be determined from both AM and PM noise contributions, increases quickly with input power as the number of stages increases. Moreover, the effective NF degrades in absolute terms significantly more in the amplifiers with lower gain. In practice, we have shown that amplifier chains with more stages can offer a more effective NF for a given output power.

We have shown that AM and PM flicker and white noise performances, in terms of single sideband noise power relative to carrier power, demonstrate comparable results at a given carrier input power independent of the number of stages, provided that amplifiers are working under small-signal conditions. However, as the number of stages is increased, the amplifiers are driven into compression at lower levels of carrier input power, and thus the noise performance degrades faster with the input power.

Finally, a general low correlation among AM and PM noises was found. However, in a particular series of amplifiers, we measured peaks in the correlation in the operating region near the 1 dB gain compression point, being more significant at some particular low-frequency offsets from the carrier frequency but also present in the white portion of the noise spectra. These kinds of peaks were also found in the derivatives of the AM–PM characteristics of those amplifiers, revealing that the non-linear mechanisms behind the noise conversion taking place inside these active devices were related to those controlling their AM to PM conversions associated with carrier signal.

**Author Contributions:** Data curation, J.P.; methodology, J.P.; software, I.B.; writing—review and editing, I.B. All authors have read and agreed to the published version of the manuscript.

**Funding:** This research was partly funded by Basque Country Government (GIUIT1104-16). Authors are part of the Electricity and Electronics Department, Faculty of Sciences and Technology, University of the Basque Country, 48940 Leioa, Spain.

**Conflicts of Interest:** The authors declare no conflict of interest.

*Electronics* **2022**, *11*, 470
11 of 11